\documentclass[a4paper, fontsize=10pt]{article}

\renewenvironment{abstract}
               {\list{}{\rightmargin\leftmargin}%
                \item[\hspace*{1cm}\small\textbf{Abstract ---}]\relax}
               {\endlist}

\usepackage[top=.8in, bottom=.8in, left=.8in, right=.8in]{geometry}
\usepackage[fleqn]{amsmath}
\usepackage{amssymb}
\usepackage{enumerate}
\usepackage{amsthm}
\usepackage[pdftex]{graphicx}
\usepackage{balance}
\usepackage{fancyhdr}
\usepackage{sidecap}
\usepackage{appendix}
\usepackage{float}
\usepackage{scrextend}
\usepackage{changepage}
\usepackage{subfigure}
\usepackage{endnotes}

\let\footnote=\endnote

\newtheorem*{Postulate*}{}

\begin{document}

\title{\textbf{On the nonreality of the PBR theorem}}

\author{Marcoen J.T.F. Cabbolet\\
        Department of Philosophy, Vrije Universiteit Brussel\\
        \small{E-mail: Marcoen.Cabbolet@vub.be}}
\date{}

\maketitle

\begin{abstract} \footnotesize The PBR theorem has been hailed as one of the most important theorems in the foundations of quantum mechanics (QM), cf. E. Samuel Reich, ``Quantum theorem shakes foundations'', Nature (2011). Here we argue that the special measurement, used by Pusey et al. to derive the theorem, is nonexisting from the Einsteinian view on QM.
\end{abstract}

\begin{SCfigure}[1.0][h!]
\centering
\includegraphics[width=0.495\textwidth]{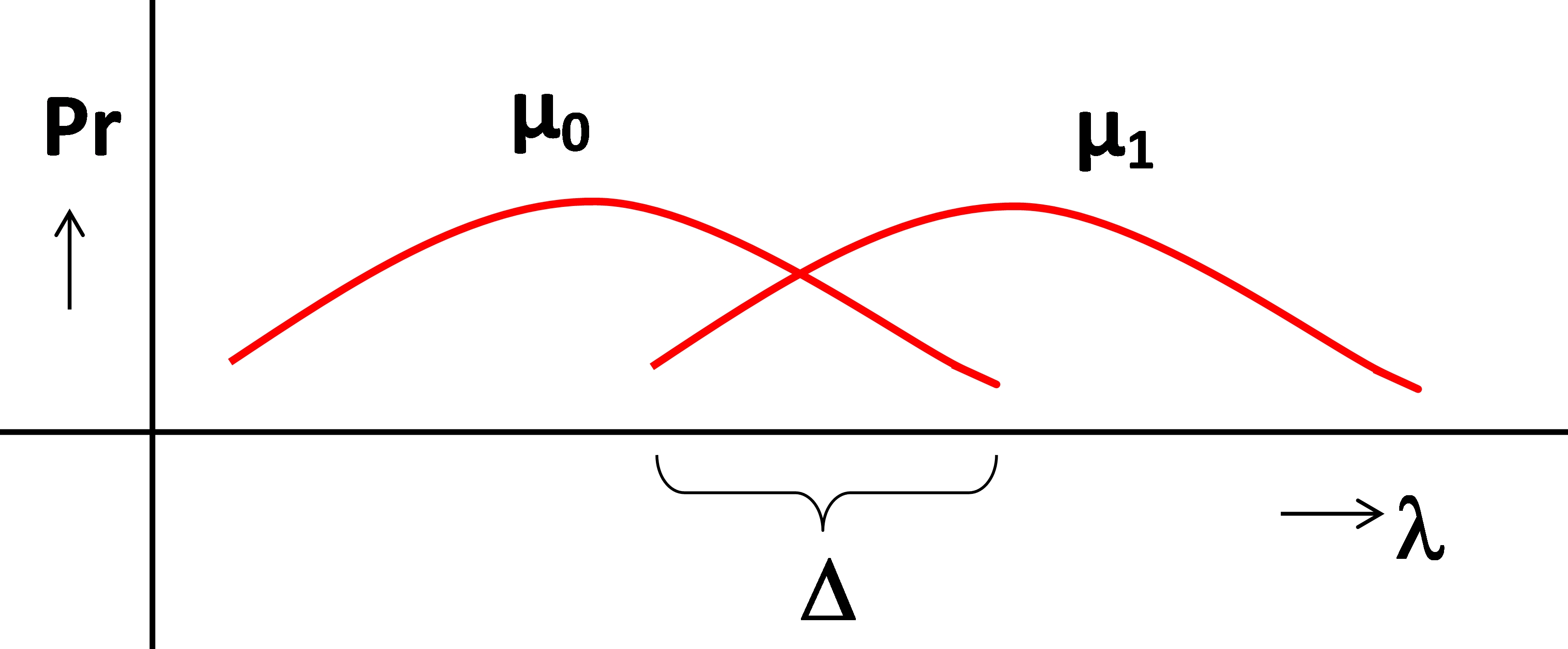}
\caption{Illustration of the probability distributions $\mu_0(\lambda)$ and $\mu_1(\lambda)$. Vertically the probability density, horizontally the possible values $\lambda$ of the physical property $\Lambda$. The probability distributions have a nonzero overlap $\Delta$.}
\label{fig:Probabilities}
\end{SCfigure}

\noindent In \cite{Pusey}, Pusey, Barrett, and Rudolph have claimed that $\psi$-epistemic quantum mechanics (QM) is inconsistent with predictions of standard QM if we
\begin{enumerate}[(i)]
  \item independently prepare two systems such that
  \begin{itemize}
    \item each system can be prepared in two different ways such that the associated quantum states $|\psi_0\rangle$ and $|\psi_1\rangle$ satisfy
{\setlength{\mathindent}{0cm}\begin{gather}
  |\ \psi_0\ \rangle = |\ 0\ \rangle \\
  |\ \psi_1\ \rangle = |\ +\ \rangle = (\ |\ 0\ \rangle + |\ 1\ \rangle)/ \sqrt{2} \\
  \langle\ \psi_0\ | \ \psi_1 \ \rangle = 1 / \sqrt{2}
\end{gather}}
where $\{ |\ 0\ \rangle, |\ 1\ \rangle\}$ is an appropriately chosen basis of the Hilbert space;
    \item the states $|\ \psi_0\ \rangle$ and $|\ \psi_1\ \rangle$ have probability distributions $\mu_0(\lambda)$ and $\mu_1(\lambda)$, respectively, for the possible values $\lambda$ of a physical property $\Lambda$, such that there is a region $\Delta$ where the support of $\mu_0(\lambda)$ and the support of $\mu_1(\lambda)$ overlap---see Fig. \ref{fig:Probabilities} for an illustration;
  \end{itemize}
  \item upon preparation of the systems, do a special joint measurement---called an ``entangled measurement'' in \cite{Pusey}---that yields a projection onto one of the following four orthogonal states, henceforth to be called `PBR states':
\begin{gather}
  |\ \xi_1\ \rangle = \frac{1}{\surd 2}\left( |\ 0\ \rangle \otimes |\ 1\ \rangle + |\ 1\ \rangle \otimes |\ 0\ \rangle\right)\\
  |\ \xi_2\ \rangle = \frac{1}{\surd 2}\left(|\ 0\ \rangle \otimes |\ -\ \rangle + |\ 1\ \rangle \otimes |\ +\ \rangle\right)\\
  |\ \xi_3\ \rangle = \frac{1}{\surd 2}\left(|\ +\ \rangle \otimes |\ 1\ \rangle + |\ -\ \rangle \otimes |\ 0\ \rangle\right)\\
  |\ \xi_4\ \rangle = \frac{1}{\surd 2}\left(|\ +\ \rangle \otimes |\ -\ \rangle + |\ -\ \rangle \otimes |\ +\ \rangle\right)
\end{gather}
where $|\ -\ \rangle = (\ |\ 0\ \rangle - |\ 1\ \rangle)/ \sqrt{2}$.
\end{enumerate}
In this letter we will argue that from the Einsteinian view on quantum mechanics, the special joint measurement proposed by Pusey et al. is nonexisting. First we show how the above PBR states can be obtained by ``entangled'' measurements on a specially prepared classical system. Thereafter we argue that no measurement exists that projects onto these states if the system under observation is prepared as in clause (i) above.
\vfill

\begin{figure}[h!]
\centering
\includegraphics[width=0.7\textwidth]{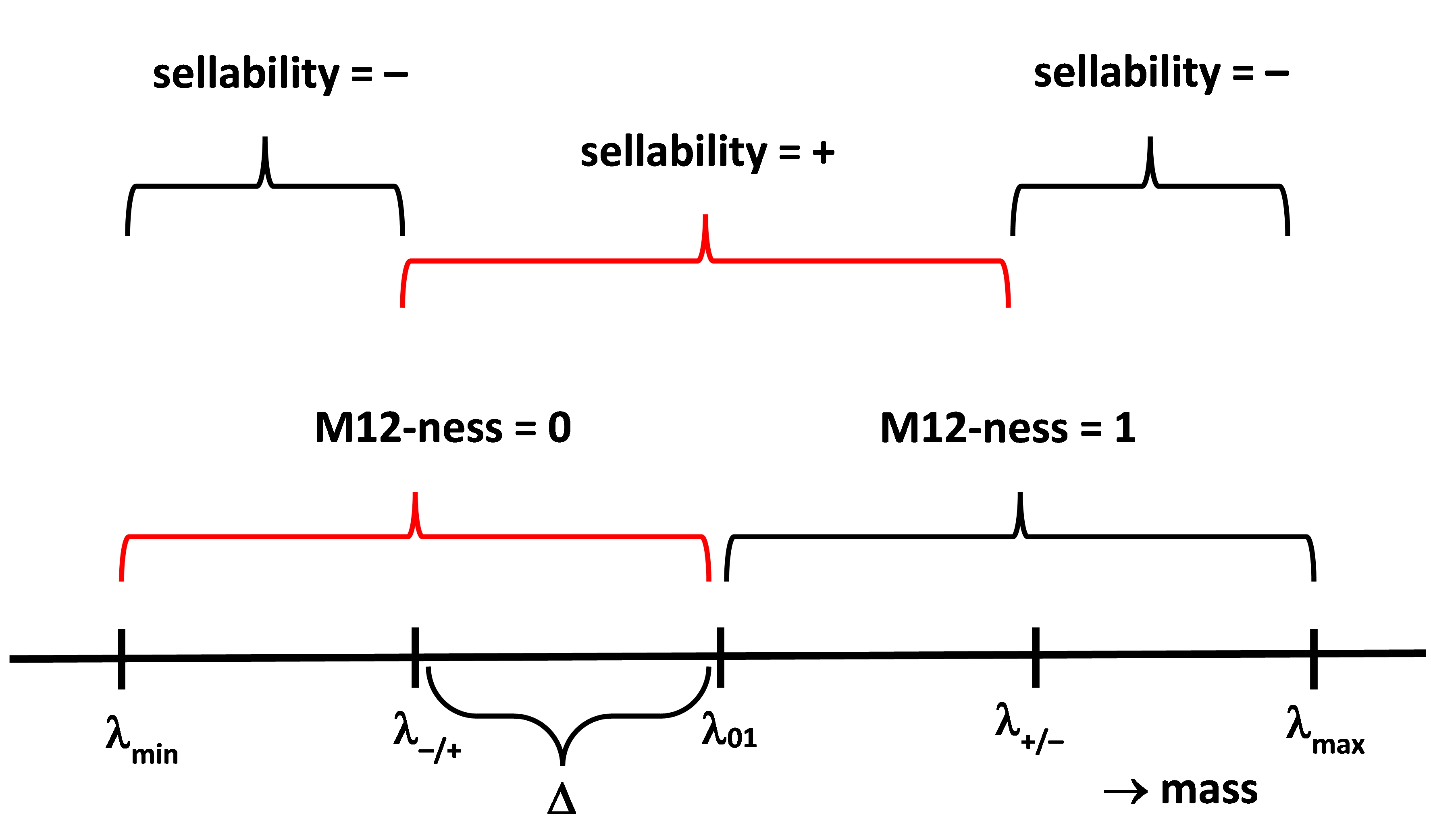}
\caption{Illustration of the masses for the various bolts. The minimum mass of any bolt is $\lambda_{\rm min}$, its maximum mass is $\lambda_{\rm max}$. If a bolt is M10, then its M12-ness is 0 and its mass is smaller than $\lambda_{01}$. If a bolt is M12, then its M12-ness is 1 and its mass is larger than $\lambda_{01}$. If a bolt is sellable, then its sellability is + and its mass is larger than $\lambda_{-/+}$ and smaller than $\lambda_{-/+}$. If a bolt is unsellable, then its sellability is -- and its mass is smaller than $\lambda_{-/+}$ or larger than $\lambda_{+/-}$. There is a interval $\Delta = (\lambda_{-/+}, \lambda_{01})$ such that bolts with a mass in $\Delta$ are both M10 and sellable.}
\label{fig:BoltsDistributions}
\end{figure}

\newpage
Consider a large enough batch of M10 bolts, and an equinumerous batch of M12 bolts. We assume that the mass distributions for M10 and M12 bolts have no overlap, so we can easily check the M12-ness of a bolt by comparing its mass $\lambda$ on a balance scale to a threshold value $\lambda_{01}$: if $\lambda > \lambda_{01}$ then we know that the bolt is M12, if $\lambda < \lambda_{01}$ then we know that the bolt is M10. If we \emph{know} that a bolt is M12, we associate to it a wave function $|\ 1\ \rangle$. If we \emph{know} that a bolt is not M12, i.e. M10, we associate to it a wave function $|\ 0\ \rangle$. If we know that a bolt is M12 then we are sure that it is not M10, and vice versa; we have
\begin{equation}\label{eq:InnerProduct10}
  \langle\ 0\ | \ 1\ \rangle = 0
\end{equation}
As it is, only the heaviest 50\% of the M10 bolts are sellable, and the lightest 50\% of the M12 bolts are sellable. We assume that the mass distributions for the sellable and unsellable bolts have no overlap, so we can easily check the sellability by checking whether its mass $\lambda$ is in between the boundary values $\lambda_{-/+}$ and $\lambda_{+/-}$ of sellability: if so, we know that the bolt is sellable; if not, we know that the bolt is unsellable. If we \emph{know} that a bolt is sellable, we associate to it a wave function $|\ +\ \rangle$ for which
\begin{equation}\label{eq:long}
  | \ +\ \rangle = \frac{1}{\sqrt{2}}\left(| \ 0\ \rangle + | \ 1\ \rangle\right)
\end{equation}
So, if we know that a bolt is sellable then the probability that it is M10 is 50\% and the probability that it is M12 is 50\%. If we \emph{know} that a bolt is unsellable, we associate to it a wave function $|\ -\ \rangle$ for which
\begin{equation}\label{eq:long}
  | \ -\ \rangle = \frac{1}{\sqrt{2}}\left(| \ 0\ \rangle - | \ 1\ \rangle\right)
\end{equation}
So, if we know that a bolt is unsellable then the probability that it is M10 is 50\% and the probability that it is M12 is 50\%. However, if we know that a bolt is sellable then the probability that it is unsellable is 0\%, and vice versa; we have
\begin{equation}\label{eq:InnerProduct10}
  \langle\ +\ | \ -\ \rangle = 0
\end{equation}
See Fig. \ref{fig:BoltsDistributions} for an illustration; comparing the red brackets in Fig. \ref{fig:BoltsDistributions} with the distributions in Fig. \ref{fig:Probabilities} shows that the probability distributions of mass are consistent with the conditions required in the PBR theorem.

An individual called Mr. Hidden is now going to prepare a batch of sets of paired envelopes---a set of paired envelopes consists of two envelopes linked by a string---by filling each set of paired envelopes in one of two ways:
\begin{enumerate}[(S1)]
  \item in one envelope is a sellable M12 bolt and in the other one an unsellable M10 bolt;
  \item in one envelope is a sellable M10 bolt and in the other one an unsellable M12 bolt.
\end{enumerate}
This is our batch: each set of paired envelopes is a system that can be measured. If one takes a system from this batch, one doesn't know what's inside the envelopes: only Mr. Hidden knows the values of the M12-ness and the sellability---we therefore call these the Hidden variables.
\newpage
For our entangled measurement, we consider four sets of measurement instructions for experimenters Alice and Bob:
\begin{enumerate}[(M1)]
  \item both Alice and Bob measure M12-ness;
  \item Alice measures M12-ness and Bob measures sellability;
  \item Alice measures sellability and Bob measures M12-ness;
  \item both Alice and Bob measure sellability.
\end{enumerate}
In each case, the measurement instruction for Alice is put in an envelop, the measurement instruction for Bob is put in an envelop, and the two envelopes are again linked by a string. We prepare a large batch of these sets of measurement instructions, such that all sets occurs in equal numbers. If one takes a set of measurement instructions from the batch, one does not know what's inside the envelopes: only by opening the envelop this becomes visible. The entangled measurement then goes as follows:
\begin{enumerate}[(i)]
  \item Alice and Bob take one system from the batch---they do not know the value of the Hidden variables;
  \item the string is cut---Alice takes one envelop, Bob the other;
  \item Alice and Bob take one set of measurement instructions from the batch;
  \item the string is cut---Alice gets one measurement instruction, Bob the other;
  \item for a measurement of M12-ness, a balance scale is used to compared its mass with $\lambda_{01}$---if $\lambda > \lambda_{01}$ then the M12-ness is 1, and if $\lambda < \lambda_{01}$ then the M12-ness is 0;
  \item for a measurement of sellability, the envelop is put on an electronic balance where an algorithm processes the signal to determine whether the mass is in the interval $I = (\lambda_{-/+}, \lambda_{+/-})$ or not---if $\lambda \in I$ then the sellability is +, if $\lambda \not\in I$ then the sellability is --.
\end{enumerate}
(See Fig. \ref{fig:BoltsDistributions} for the meaning of the $\lambda$'s.) Note that a measurement of sellability reveals no information about the M12-ness, and that a measurement of M12-ness reveals no information about the sellability.

It is then easy to show that, with the values of the Hidden variables as in S1 and S2 above, \emph{over time} the four PBR states are obtained from \emph{a series of measurements}. So for a \emph{single measurement}, we \emph{know} that M1 yields $|\ \xi_1\ \rangle$, M2 yields $|\ \xi_2\ \rangle$, M3 yields $|\ \xi_3\ \rangle$, and M4 yields $|\ \xi_4\ \rangle$. There are countless variations of the above experiment that give the same result.\\
\ \\
Having shown how the four PBR states \emph{can} be obtained from measurements on a specially prepared classical system, we now show that these states \emph{cannot} be obtained from measurements on two combined systems that have been independently prepared as required by the PBR-theorem. For starters, when we talk about a measurement in quantum mechanics we \emph{always} have in mind a measurement of a value of a property of a system. And here, there are two properties of the system involved:
\begin{enumerate}[(i)]
  \item A-ness, with possible values $a_1$ and $a_2$---if we know that the system has an A-ness of $a_1$ then we associate to it the wave function $|\ 0\ \rangle$, and if we know that the system has an A-ness of $a_2$ then we associate to it the wave function $|\ 1\ \rangle$;
  \item P-ness, with possible values $p_1$ and $p_2$---if we know that the system has a P-ness of $p_1$ then we associate to it the wave function $|\ +\ \rangle$, and if we know that the system has a P-ness of $p_2$ then we associate to it the wave function $|\ -\ \rangle$.
\end{enumerate}
We then assume that these state vectors satisfy all conditions of the PBR theorem. To obtain the PBR states, we must clearly be able to measure the A-ness and the P-ness of each of the two combined systems, because
\begin{itemize}
  \item to obtain the state $|\ \xi_1\ \rangle$ we must measure the A-ness of both systems;
  \item to obtain the state $|\ \xi_2\ \rangle$ we must measure the A-ness of the $1^{\rm st}$ system and the P-ness of the $2^{\rm nd}$ system;
  \item to obtain the state $|\ \xi_3\ \rangle$ we must measure the P-ness of the $1^{\rm st}$ system and the A-ness of the $2^{\rm nd}$ system;
  \item to obtain the state $|\ \xi_4\ \rangle$ we must measure the P-ness of both systems.
\end{itemize}
\nopagebreak
The crux is now that in order for these measurements of A-ness or P-ness to have an outcome, the Einsteinian point of view is that the system \emph{already had} both an A-ness and a P-ness prior to measurement---the orthodox view on quantum mechanics, on the other hand, is that a system prepared with an A-ness \emph{does not} have a P-ness prior to a measurement of the P-ness and vice versa, cf. \cite{Cabbolet}. So, from the Einsteinian point of view it cannot be the case that a system prepared with an A-ness does not have a P-ness, because then a measurement of the P-ness would yield no value---then the measurement would not project onto any state vector.
\newpage
\noindent Elaborating, applying the Born rule yields the equation
\begin{equation}\label{eq:0andPnessP1}
  {\rm P}^{|\ 0\ \rangle}(p_1) = (\langle\ +\ |\ 0\ \rangle)^2 = 1/2
\end{equation}
meaning that the probability of finding the value $p_1$ upon measuring the P-ness of a system with associated quantum state $|\ 0\ \rangle$ is 50\%. So in half of the cases, a measurement of the P-ness yields the value $p_1$. But that means that in those cases, the P-ness was already $p_1$ prior to measurement. Likewise, the equation
\begin{equation}\label{eq:0andPnessP2}
  {\rm P}^{|\ 0\ \rangle}(p_2) = (\langle\ -\ |\ 0\ \rangle)^2 = 1/2
\end{equation}
means that the probability of finding the value $p_2$ upon measuring the P-ness of a system with associated quantum state $|\ 0\ \rangle$ is 50\%. But in the cases where we find the value $p_2$ upon measurement, the P-ness was already $p_2$ prior to measurement. So if we have associated the quantum state $|\ 0\ \rangle$ to a system, then its P-ness has a definite value, although we do not know which value that is if we only know that the A-ness is $a_1$. By the same token, the equations
\begin{gather}
  {\rm P}^{|\ +\ \rangle}(a_1)= (\langle\ 0\ |\ +\ \rangle)^2 = 1/2\\
  {\rm P}^{|\ +\ \rangle}(a_2)= (\langle\ 1\ |\ +\ \rangle)^2 = 1/2
\end{gather}
mean that the probability of finding the value $a_1$ c.q. $a_2$ upon measuring the A-ness of a system with associated quantum state $|\ +\ \rangle$ is 50\%. So likewise, if we have associated the quantum state $|\ +\ \rangle$ to a system, then its A-ness has a definite value---this value is $a_1$ in half of the cases and $a_2$ in the other half of the cases---but we do not know which value that is if we only know that the P-ness is +.

But that means that if we combine two systems---each prepared in the state $|\ 0\ \rangle$ or $|\ +\ \rangle$---and we measure the A-ness of the first system and the P-ness of the second system, then it is not the case that we find half the time $|\ 1\ \rangle\otimes|\ +\ \rangle$ and half the time $|\ 0\ \rangle\otimes|\ -\ \rangle$. We also find $|\ 1\ \rangle\otimes|\ -\ \rangle$ and $|\ 0\ \rangle\otimes|\ +\ \rangle$. The systems are prepared independently, so it is not possible that
\begin{enumerate}[(i)]
  \item if we find the A-ness of the first system to be $a_1$, then the P-ness of the second system can only be $p_2$;
  \item if we find the A-ness of the first system to be $a_2$, then the P-ness of the second system can only be $p_1$;
\end{enumerate}
In other words, the projection on $|\ \xi_2\ \rangle$ is not possible. Similarly for $|\ \xi_3\ \rangle$, which involves a measurement of the P-ness of the first system and the A-ness of the second system---projection on $|\ \xi_3\ \rangle$ is not possible.

Moreover, Eqs. \eqref{eq:0andPnessP1} and \eqref{eq:0andPnessP2} mean that if we prepare a system in the state $|\ 0\ \rangle$, then half the time its P-ness is $p_1$ and half the time its P-ness is $p_2$. We can write this down as
\begin{equation}
|\ 0\ \rangle = \left( |\ 0, +\ \rangle + |\ 0, -\ \rangle \right) / \surd 2
\end{equation}
The state vector $|\ 0,+\ \rangle$ satisfies
\begin{gather}
\langle\ 0\ |\ 0,+\ \rangle =  \langle\ +\ |\ 0,+\ \rangle = 1\label{eq:AP1}\\
\langle\ 1\ |\ 0,+\ \rangle =  \langle\ -\ |\ 0,+\ \rangle = 0\label{eq:AP0}
\end{gather}
meaning that if we associate to the system the state vector $|\ 0,+\ \rangle$ then we \emph{know} that its A-ness is $a_1$ and its P-ness is $p_1$: Eq. \eqref{eq:AP1} expresses that the probability that a measurement of the A-ness c.q. the P-ness gives $a_1$ c.q. $p_1$ is then 100\%, and Eq. \eqref{eq:AP0} expresses that the probability that a measurement of the A-ness c.q. the P-ness gives $a_2$ c.q. $p_2$ is then 0\%. Likewise for the state vector $|\ 0, -\ \rangle$. Thus speaking, if we prepare the two systems both in the state $|\ 0\ \rangle$, then we have
\begin{equation}
|\ 0\ \rangle\otimes|\ 0\ \rangle = \left( |\ 0, +\ \rangle\otimes|\ 0, +\ \rangle +  |\ 0, +\ \rangle\otimes|\ 0, -\ \rangle + |\ 0, -\ \rangle\otimes|\ 0, +\ \rangle + |\ 0, -\ \rangle\otimes|\ 0, -\ \rangle \right) / 2
\end{equation}
Likewise for other combinations of $|\ 0\ \rangle$ and $|\ +\ \rangle$. However, writing $|\ 00, ++\ \rangle$ for $|\ 0, +\ \rangle\otimes|\ 0, +\ \rangle$, we have
\begin{equation}\label{eq:zero}
  \langle\ \xi_1\ |\ 00, ++\ \rangle = \langle\ \xi_2\ |\ 00, ++\ \rangle = \langle\ \xi_3\ |\ 00, ++\ \rangle = \langle\ \xi_4\ |\ 00, ++\ \rangle = 0
\end{equation}
for each of the four PBR vectors $|\ \xi_j\ \rangle$. That means that in 25\% of all cases, the measurement on the two combined systems, each prepared in the state $|\ 0\ \rangle$ or $|\ +\ \rangle$, does not yield a projection onto any of the PBR vectors $|\ \xi_j\ \rangle$, contrary to the assumption by Pusey et al. that the measurement (always) yields a projection onto one of the four state vectors $|\ \xi_j\ \rangle$.

Alternatively, we can say that the ``entangled measurement'' that projects onto one of the PBR states is not a combination of measurements of A-ness or P-ness but a measurement of a property of the combined systems, represented by an operator $\hat{\Xi}$, whose spectrum has four eigenvalues with corresponding eigenvectors $|\ \xi_j\ \rangle$. But then we have tacitly assumed that in all cases the combination of the two systems, each prepared in the state $|\ 0\ \rangle$ or $|\ +\ \rangle$, actually has that property represented by $\hat{\Xi}$. The foregoing shows that this assumption is false from the Einsteinian point of view on quantum mechanics---in particular, Eq. \eqref{eq:zero} shows that in 25\% of the cases the combined systems do not have the property represented by $\hat{\Xi}$.
\newpage

Summarizing, the validity of the PBR-theorem leans on the tacit assumption that the ``entangled measurement'', which projects on the states $|\ \xi_j\ \rangle$, exists. The argument against the PBR theorem here is that this ``entangled measurement'' is nonexisting from the Einsteinian point of view on quantum mechanics. Since the PBR theorem puts no further restrictions on the nature of the two systems, the predicted failure to project onto the PBR states in 25\% of the cases can be experimentally verified for the price of a few lollipops by children doing simple measurements on bolts in envelopes.

\end{document}